\documentclass{article}

\usepackage{arxiv}

\usepackage[utf8]{inputenc} 
\usepackage[T1]{fontenc}    
\usepackage{hyperref}       
\usepackage{url}            
\usepackage{booktabs}       
\usepackage{amsfonts}       
\usepackage{nicefrac}       
\usepackage{microtype}      
\usepackage{lipsum}

\title{Reducing malicious use of synthetic media research: 
\\ Considerations and potential release practices for machine learning} 

\author{
  Aviv Ovadya \\
  The Thoughtful Technology Project\\
  \texttt{aviv.ovadya@thoughtfultech.org} \\
   \And
 Jess Whittlestone \\
  Leverhulme Centre for the Future of Intelligence\\
  University of Cambridge\\
  \texttt{jlw84@cam.ac.uk} \\
}

\begin{document}
\maketitle

\begin{abstract}
The aim of this paper is to facilitate nuanced discussion around research norms and practices to mitigate the harmful impacts of advances in machine learning (ML). We focus particularly on the use of ML to create ``synthetic media'' (e.g. to generate or manipulate audio, video, images, and text), and the question of what publication and release processes around such research might look like, though many of the considerations discussed will apply to ML research more broadly. We are not arguing for any specific approach on when or how research should be distributed, but instead try to lay out some useful tools, analogies, and options for thinking about these issues.
\vspace{.6\baselineskip}

We begin with some background on the idea that ML research might be misused in harmful ways, and why advances in synthetic media, in particular, are raising concerns. We then outline in more detail some of the different paths to harm from ML research, before reviewing research risk mitigation strategies in other fields and identifying components that seem most worth emulating in the ML and synthetic media research communities. Next, we outline some important dimensions of disagreement on these issues which risk polarizing conversations.
\vspace{.6\baselineskip}

Finally, we conclude with recommendations, suggesting that the machine learning community might benefit from: working with subject matter experts to increase understanding of the risk landscape and possible mitigation strategies; building a community and norms around understanding the impacts of ML research, e.g. through regular workshops at major conferences; and establishing institutions and systems to support release practices that would otherwise be onerous and error-prone.

\end{abstract}


\section{Introduction}

Technological advances can result in harm to both individuals and societal structures, through accidents, unintended consequences, and malicious use -- even as the same advances provide incredible benefits. Concern about harms resulting from advances in machine learning (ML) have risen dramatically in recent years, with a growing community of researchers and practitioners doing crucial work to address issues such as the fairness, accountability, and transparency of deployed systems \cite{whittlestone2019}. In this paper, we focus primarily on the ways that advances in ML research might be deliberately misused by malicious actors (which we refer to as ``mal-use'') \cite{brundage2018malicious}, though we also touch on other unintended consequences.\footnote{Unintended consequences are also particularly salient for the case of synthetic video and audio, where the simple existence of the technology, regardless of its use, can allow bad actors to claim that evidence of e.g. corruption or war crimes were synthesized in order to avoid accountability. For example, allegations of videos being faked have been used to justify a coup in Gabon \cite{breland2019}, and exculpate a cabinet minister in Malaysia \cite{ker2019}. 
}

One specific example of where mal-use concerns have been raised recently is around the use of ML in \textit{synthetic media}\footnote{Not all synthetic media techniques involve ML. For example, traditional computer graphics techniques are also used for image and video manipulation. For simplicity of language, we will not consider these distinctly as the considerations for research practices are very similar.}: the manipulation and generation of increasingly realistic audio, video, images, and text. It is now possible to produce synthetic images of faces that are almost indistinguishable from photographs \cite{karras2019}. It is becoming easier to manipulate existing videos of people: replacing a person’s facial expressions or movements with those of another person \cite{chan2018, thies2016}, and/or overlaying synthesized speech which imitates a person’s voice \cite{skerry2018towards}. Language models have advanced to the point where, given a sentence prompt, they can write articles that appear at least superficially convincing \cite{radford2019}. These advances could be used to impersonate people, sway public opinion, or more generally spread doubt about the veracity of all media. Modern synthetic media is in fact already being used for harm: face-swapping tools are being used to harass journalists \cite{harwell2018}, synthetic voices are being used for financial crimes \cite{bbc2019}, and synthetic faces have allegedly been used for espionage \cite{satter2019}. 

These examples bring into focus the need for humility around what we don’t know concerning potential impacts of technology, and perhaps suggests that we must develop better systems for understanding such impacts. Related concerns have sparked debate about responsible research practices in ML \cite{pai2019}, and in particular whether sometimes it is appropriate to withhold open publication of some aspects of this research. 


We begin with some background on the idea that ML research might be misused in harmful ways, and why advances in synthetic media, in particular, are raising concerns. We then review research risk mitigation strategies in other fields and identifying components that may be worth emulating in the ML and synthetic media research communities. Finally, we outline some important dimensions of disagreement on these issues which risk polarizing conversations, before concluding with some recommendations for research and practice.

\section{How research can lead to harm}
\label{sec:1}

How might advances in ML and synthetic media end up resulting in harm? We begin by spelling this out in more detail: different ways that research might empower malicious actors, and some possible paths to harm from this. 

\subsection{Types of hazard}

We can distinguish several different types of “information hazard” that might result from machine learning research in general. \footnote{This is based loosely on a taxonomy from \cite{bostrom2011n}
}

\begin{itemize}
    \item \textbf{Product hazard:} \textit{Research produces software that can be directly used for harm} (e.g. rootkits for computer hacking). Product hazards increase the likelihood of mal-use by adversaries with minimal technical capabilities (e.g. `script kiddies' who use existing programs or scripts to hack into computers but lack the ability to write their own, or less sophisticated information warfare operations), or those who have weak motivations for harm (e.g. online hacking or deception just `for fun').
    \item \textbf{Data hazard:} \textit{research produces detailed information or outputs which if disseminated, create risk of use for harm} (e.g. easy nuclear blueprints; models, training data or code for harmful software). Data hazards increase the likelihood of mal-use by adversaries with some technical capabilities, but without e.g. access to high-quality researchers.
    \item \textbf{Attention hazard:} \textit{research which directs attention towards an idea or data that increases risk}  (e.g. the idea that it is possible to use voice-cloning for phishing). Attention hazards increase the likelihood of mal-use by adversaries who may not have realized that their objectives can be aided by new technologies.
\end{itemize}

An important thing to note is that potential mitigations will be \textit{different for each type of hazard -- and potentially even in conflict}. One way to mitigate attention hazards is to be very careful about talking to media organizations and others with large reach about ways that research advances could be used maliciously (such as in voice cloning, for example). At the same time, raising wider concern about malicious use cases of ML progress might be exactly what is needed to incentivize mitigations for data hazards or product hazards (e.g. some public concern may be required to ensure that tech platforms prioritize developing technology to identify voice cloning, or for telecoms to build mitigating infrastructure to make phone number spoofing harder). 

\subsection{The path to harm}

But how might these different types of  `hazard' actually lead to real-world harms? It is worth connecting the dots here between the theoretical potential for mal-use, and what actually makes significant harm more likely. 

Below we talk through some factors influencing whether a capability leads to sustained mal-use in practice. We use artificial voice cloning as an illustrative example, as a relatively new capability with many useful applications (e.g. in voice translation and audio editing) but also significant potential for mal-use (e.g. in scams, political propaganda, and market manipulation).

\begin{enumerate}
    \item \textit{ \textbf{Awareness}: Do actors with malicious intent know about a capability and believe it can help them?}
    \vspace{.3\baselineskip}
    
    We can break this down into:
    
    \begin{itemize}
        \item \textbf{Attention of adversaries}: Are malicious actors likely to realize that they could use a new capability to further their ends? If adversary groups are already using closely related methods, this is much more likely: for example, if edited voice clips are already being used for political manipulation, groups doing this are more likely to pay attention to demonstrations of voice cloning.
        \item \textbf{‘Convincibility’ of those with resources}: Are there compelling arguments, perhaps by authoritative third parties, for the effectiveness of new capabilities? For example, a scammer who realizes that voice cloning is useful might need to be able to convince a superior that this technology is effective enough to justify the costs and overcome institutional inertia.
    \end{itemize} 
    \vspace{.5\baselineskip}

    \item \textit{ \textbf{Deployment}: How difficult is it for adversaries to weaponize this capability in practice?}
    \vspace{.3\baselineskip}
    
    For a capability to be deployed for malicious purposes, adversaries not only need to be aware but to have the necessary skills and resources to productize and weaponize the capability. This isn’t a binary -- e.g. having ML expertise vs. not -- but rather many different factors will influence how easy a capability is to weaponize. At the extreme, we might have a product which can be immediately used by anyone, regardless of technical capability (such as free to use voice cloning software).
    \vspace{.3\baselineskip}
    
    Factors that influence the ease of deployment for mal-use include:
    
    \begin{itemize}
        \item \textbf{Talent pipelines}: How difficult is it to source someone who can apply a new capability for the desired use case? (e.g. do malicious actors need someone with machine learning experience, programming experience, or can they just use a program directly to achieve their goals?) \cite{satter2019}.
        \item \textbf{Reproducibility}: How difficult is it to reproduce a capability given the information available? (e.g. is it easy to replicate a voice cloning capability given the available papers, models, code, etc.?)
        \item \textbf{Modifiability}: How difficult is it to modify or use a system in order to enable mal-use? (e.g. if a voice cloning product makes it difficult to clone a voice without consent or watermarks, how hard is it to overcome those limitations?) 
        \item \textbf{Slottability}: Can new capabilities be slotted into existing organizational processes or technical systems? (e.g. are there already established processes for phone scams into which new voice generation capabilities can be slotted easily, without any need to change goals or strategy?)
        \item \textbf{Environmental factors}: How does the existing `environment' or `infrastructure’ impact the usefulness of the new capability for malicious actors? (E.g. currently, in the US it is easy to `spoof' phone numbers to make it appear like a call is coming from a family member, which could impact the likelihood of voice cloning being weaponized for phone scams.)
    \end{itemize}
    
    Websites now enabling anyone to instantly generate seemingly photorealistic faces are a concrete example of deployment barriers falling away and making mal-use easier. It had been possible for well over a year to generate synthetic images of faces with fairly high quality, but such websites have enabled anyone to do so with no technical expertise. This capability can also immediately slot into existing processes, such as fake account creation. Previously, malicious actors would often use existing photos of real people, which could be identified with reverse image search \cite{goth2017}, unlike wholly generated synthetic images.
    \vspace{.5\baselineskip}
    
    \item \textit{ \textbf{Sustained use}: How likely is it that a capability will lead to sustained use with substantial negative impacts?}
    \vspace{.2\baselineskip}
    
    Even if adversaries are aware of and able to weaponize some new capability, whether or not this leads to sustained use depends on:
    
    \begin{itemize}
        \item \textbf{Actual ROI}: If malicious actors believe that the return on investment (ROI) for using a capability is low they might not continue to use it in practice. For example, if a form of mal-use is easy to detect, then adversaries might decide it’s not worth the risk or might be shut down very quickly.
        \item \textbf{Assessment of ROI}: If malicious actors have no way of assessing whether new capabilities are helping them better achieve their goals, or if their assessments are flawed, they might not continue to put resources into using those capabilities.
    \end{itemize}
\end{enumerate}

\subsection{Access ratchets}

We can think of this as a kind of progression, from a theoretical capability to scaled-up use in practice. Once a technology has progressed down this path and has become easy to use, and proven to have high ROI for mal-use, it can be much more difficult to address than at earlier stages -- we call this the \textbf{access ratchet} (like a ratchet, increased access to technology cannot generally be undone). For any capability with potential for mal-use, it is therefore worth thinking about where it currently sits on this progression: how much attention and interest it is receiving; whether it has been weaponized and/or how costly it would be to do so; and whether it’s likely to be, or already in sustained use. This can help us think more clearly about where the greatest risks of mal-use are, and different kinds of interventions that might be appropriate or necessary in a given situation.

Researchers may argue that a capability is unlikely to cause harm since it has not been used maliciously yet. What this doesn’t address is the fact that a capability which has not yet been used maliciously might sit anywhere along this progression, which makes a huge difference to how likely it is to cause harm. For example, Face2Face, a technique for real-time facial reenactment (i.e. changing a person’s expressions in a video), has existed for 3 years but not been developed into any products that can easily be used. This lack of productization makes harmful use vastly less likely, especially given the competition for AI and engineering talent today. It is also worth considering how costly it would be to make a given capability easier to misuse: even the DeepFake application, which is more accessible to non-technical users, is currently resource-intensive to weaponize in practice.

\subsection{Indirect harms}

Sometimes the path to harm from synthetic media research will be fairly direct and immediate: such as a person losing their money, returning to our example of voice cloning being used in financial scams.

But in other cases, improved synthetic media capabilities might cause harm in more complex and indirect ways. Consider the case where misinformation purveyors get hold of sophisticated synthetic media capabilities and use them to win substantial democratic power, which they then use to control narratives further and undermine any mitigation efforts (not an uncommon path from democracy to authoritarianism). We can think about this as a \textbf{disinformation ratchet}: the ability to use disinformation to enhance one’s ability to distribute further disinformation; and the opportunity for this type of ratchet can be influenced by new technology impacting media distribution channels and capabilities.

These less direct kinds of harms may be harder to anticipate or imagine, but in the long-run may be much more important -- particularly if they influence the future development of technology in ways that undermine our ability to deal with future threats. We suggest that it’s particularly important to consider these kinds of “sociotechnical-path dependencies” as well as more direct and immediate threats, and what kinds of risk mitigation strategies might best address them. 

\section{Mitigating harm through release practices}

There is unlikely to be any `one size fits all' solution to mitigating mal-use of ML research: the path to harm will look very different across contexts, and potential harms need to be weighed against benefits which will also vary depending on the area. We therefore need discussion about different approaches to mitigating mal-use: including around what research is conducted in the first place; standards and procedures for risk assessment; and processes for deciding when and how to release different types of research outputs. Here we focus particularly on the latter -- how careful release practices might help mitigate mal-use within ML research. 

However, this is not to suggest we think release practices are the main or even necessarily the most important component of mitigating mal-use. Another crucial piece is how research directions are chosen and prioritized in the first place. This is challenging because much of ML research often involves developing general capabilities which can then be applied to a variety of different purposes -- we can’t simply decide to build only `beneficial' ML capabilities. That aside, we still may be able to say some very general things about the kinds of capabilities that are more likely to be broadly beneficial, or the kinds of problems that should ideally be driving ML research. It is also important to think about what types of research are encouraged/discouraged by conferences, journals, funders, job interviewers and so on. 

\subsection{Challenges to mitigating harm}

First, it's worth considering some of the serious challenges to attempting to decrease harm by limiting access to research:

\begin{itemize}
    \item The \textbf{composition problem}: Two independent pieces of research that seem innocent can be combined in ways that enable significant malicious use.\footnote{This might be particularly challenging given the success of transfer learning.}
    \item The \textbf{slow drip problem}: Research advancement can be a slow and continuous evolution, where it’s difficult to draw the line between research that is dangerous and that which is not.
    \item The \textbf{conflation problem}: Many of the underlying goals of various fields of research (natural language processing, computation photography, etc.) may be directly weaponizable if achieved. For example, the ability to create convincing dialogue can be used to both support people or manipulate people at scale.
    \item The \textbf{defector problem}:  Even if researchers in some regions or organizations cooperatively decide not to pursue or publish a particular area of research, those agreements might not be followed by “defectors” who then gain a competitive edge.
\end{itemize}

These challenges may seem daunting even for those who would advocate for limiting release of some forms of research. They also motivate the development of a nuanced menu of options for release practices, and careful evaluation of the efficacy of whatever measures are chosen. Even without overcoming these challenges, it is possible that release practices could substantially mitigate harm if they impact the rate of deployment of mal-use technology.\footnote{In terms of the ratchet terminology used earlier, delaying release of research could slow down the speed of an `access ratchet' (i.e. slowing down widespread access to a technology), potentially providing enough extra time strengthen a `sociotechnical immune system' that could halt a disinformation ratchet.}

\subsection{A brief tour of analogs in other fields}

There is precedent in several other fields of research -- including biotechnology and information security -- for establishing processes for reducing the negative risks of research and release. A good first step would, therefore, be to look at what we can learn from these fields for the case of ML research. Here we present some promising practices identified from other fields.

A caveat: just because research norms and processes exist in other fields, it does not necessarily mean that they are widely and coherently used in those fields, or that they provide a net positive impact. Evaluating which research practices have been adopted and work well across different fields is out of scope for this short paper, but would certainly be valuable to look into further. 

\subsubsection{Biosafety}

Biosafety processes and principles exist to ensure safe handling of infective microorganisms in biology/biotechnology research \cite{who2004}. Some key components of biosafety practices include:
\begin{itemize}
    \item \textit{Procedures}: Steps and rules that must be followed, e.g. for decontamination (including basics such as wearing gloves and shoes).
    \item \textit{Lab safety officer}: An internal role responsible for enforcing safety.
    \item \textit{Training}: Learning safety processes via peers/programs.
    \item \textit{Architecture}: Incorporating safety considerations into building and tool design (e.g. the design of doors and airflow).
    \item \textit{Audits}: Providing external accountability, usually at random times via local government.
    \item \textit{Safety level designations}: Different microorganisms classified by risk group (e.g. Ebola is level 4) with different safety procedures for different levels (e.g. level 1 is open bench work, level 4 requires special clothing, airlock entry, special waste disposal, etc.).
    \item \textit{Safety level definers}: Organisations who determine safety levels, e.g. the Centers for Disease Control and Prevention (CDC).
\end{itemize}

\subsubsection{Computer/Information security}

Various practices exist in the field of information security to prevent exploitation of vulnerabilities in important systems. Key components include:

\begin{itemize}
    \item \textit{OPSEC (‘operations security’)}: Procedures for identifying and protecting critical information that could be used by adversaries. Includes identification of critical information, analysis of threats, vulnerabilities, and risks, and application of appropriate measures.
    \item \textit{Architecture}: Use systems that are “secure by design” and so keep you secure automatically where possible.
    \item \textit{Coordinated/responsible disclosure}: Processes to ensure that exploits which could affect important systems are not publicly disclosed until there has been an opportunity to fix the vulnerability.
    \item \textit{ISACs/CERTs} (Information Sharing \& Analysis Centers/Computer Emergency Response Teams): Disclosure coordination entities.
\end{itemize}

\subsubsection{Institutional Review Boards (IRBs)}

IRBs are designed to protect human subjects in biomedical and behavioral research (including e.g. clinical trials of new drugs or devices and psychology studies of behavior, opinions or attitudes) \cite{enfield2008}: 

\begin{itemize}
    \item \textit{Case dependent scrutiny}: Research proposals are assessed on a case-by-case basis using external expert evaluation, and are determined to either be: (a) exempt (when risks are minimal), (b) expedited (slightly more than minimal risk), or (c) full review (all other proposals).
    \item \textit{Approval rubrics}: Criteria for approval of research proposals include: having sound research principles to minimize risk to subjects; establishing that risks to subjects are reasonable relative to anticipated benefits; selecting subjects in equitable ways, and avoiding undue emphasis on a vulnerable population.
    \item \textit{External expert \& community evaluation}: Studies are reviewed by people who have expertise in the research and in the impacts of the work (such as community members).
    \item \textit{Continuous evaluation}: Process can be ongoing, not one-time, with periodic updates: the IRB can suspend or terminate previously approved research.
\end{itemize}

This is not meant to be exhaustive but demonstrates a variety of systems that have been used to mitigate negative risks of research and release. Other analogs worth exploring include those around nuclear technology, spam detection, classified information, and environmental impact.

\subsection{Potential release practices}

What should ML and synthetic media research emulate from these other fields? Many aspects of these practices and processes may be applicable in ML research, including particularly: external expert evaluation of risks and appropriate responses, coordinated/responsible disclosure, training in responsible research processes, disclosure coordination entities, safety level designations, safety level defining entities, and case-dependent response (depending on safety levels).

Reframing and renaming all of these practices and processes to focus on the release of potentially hazardous ML systems leaves us with the following components that may be needed in ML:

\begin{itemize}
    \item \textbf{Release options}: Different options for release.
    \item \textbf{Release rubric}: Guidelines for when to use each type (decided by case-dependent evaluation).
    \item \textbf{Release rubric processes}: How to do case-dependent evaluation.
    \item \textbf{Release coordination}: Who decides/gets access.
    \item \textbf{Release training}: How to learn processes/norms.
    \item \textbf{Release process entities}: Who manages all of this?
\end{itemize}

Each of these components can be broken down further; we explore “release options” here as an example. 

\subsection{Release options}

The question of how to release research with potential for mal-use is not a binary one: there are many different choices to make beyond simply `release' or `don’t release'. Focusing on this binary choice can lead the debate around openness of ML research to become very polarized.

Some important dimensions we might consider when thinking about release strategies include:

\begin{itemize}
    \item \textbf{Content: \emph{What is released}}
    
    Potential options include: 
    \vspace{-.35\baselineskip}
    \begin{itemize}
        \item A fully runnable system (with varying power)
        \item A modifiable system (with varying modifiability)
        \item Source code (varying versions)
        \item Training data (varying sizes)
        \item Trained models (varying strength/fine-tunability/data-needs)
        \item Paper/concept (varying detail level)
        \item Harmful use case ideas (varying detail level)
    \end{itemize} 
    \item \textbf{Timing: \textit{When it is released}}
    
    Potential options include: 
    \vspace{-.35\baselineskip}
    \begin{itemize}
        \item Immediate release 
        \item Timed release: Set a specific time to release components, allowing time for mitigation of any potential harms. This is common in information security.
        \item Periodic evaluation: Don’t release immediately, but set a time period/intervals (e.g. every 2 months), at which point an evaluation is done to reassess the risk of release given mitigation progress.
        \item Evented release: Wait to release until some particular type of external event (e.g. someone else replicating or publicizing the same technology).
        \item Staged release: Release systems of successively increasing levels of power on a fixed timeline, or triggered by external events.
    \end{itemize}
    
    \item \textbf{Distribution: \textit{Where/Who it is released to}}
    
    Potential options include: 
    \vspace{-.35\baselineskip}
    \begin{itemize}
        \item Public access (with varying degrees of publicity)
        \item Ask for access: Anyone who wants access to data or a system asks and is approved on a case-by-case basis, potentially with specific requirements around use. 
        \item Release safety levels: People and possibly organizations can request to be recognized as ‘safe’, after auditing and approval they gain the ability to access all material at a given safety level.
        \item Access communities: Research groups developing their own trusted communities through informal processes which all have access to shared repositories.
    \end{itemize}
\end{itemize}

Within the domain of synthetic media, it’s worth diving deeper into potential mitigations specific to products, models, and demos relevant to that space.
 
There are a number of mechanisms researchers and companies can use to reduce malicious use from general synthetic media systems that allow e.g. virtual impersonation:

\begin{itemize}
    \item \textbf{Consent}: Requiring consent by those being impersonated.
    \item \textbf{Detectability}: Intentionally not trying to thwart detection.
    \item \textbf{Watermarking}: Embedding context about modifications/original.
    \item \textbf{Referenceability}: Centrally storing all modifications for reference.
\end{itemize}

It's important to note that none of these are perfect -- they are part of a ``defense in depth''. It is also possible to add constraints on synthesis (e.g. ensuring that only particular faces can be generated through the system).

\subsection{Examples in practice}

This menu of options is valuable in theory, but it’s also worth briefly exploring some examples in practice. One of the most notable public positions in this space comes from Google: ``We generally seek to share Google research to contribute to growing the wider AI ecosystem. However we do not make it available without first reviewing the potential risks for abuse. Although each review is content specific, key factors that we consider in making this judgment include: risk and scale of benefit vs downside, nature and uniqueness, and mitigation options.'' \cite{google}

Beyond Google, a number of labs have had to consider these issues. As mentioned earlier, the researchers behind Face2Face and those behind many other synthetic media systems have chosen not to share their code. Some researchers have released code but intentionally made it difficult to use for non-experts.
 
Different product companies in this space are also exploring mitigations.\footnote{For more on this, see this crowdsourced list of organizations and their actions to mitigate risk: http://bit.ly/synth-ethics.} Synthesia is only working with closely vetted clients. Lyrebird, which enables voice cloning, makes it more difficult to impersonate someone without consent by requiring users to speak particular phrases instead of just training on arbitrary provided data.

\section{Disagreements around release practices}

Different people and groups will have differing views on which kinds of release strategies should be used when. Here we lay out some different dimensions on which people may disagree which affect their views about release strategies for ML research. Our aim is to recognize that genuine divides exist and can lead to polarization of opinion, but that more nuanced discussion can prevent this.

\subsection{Value trade-offs}

Some disagreements stem from fundamental views about the value of openness vs. caution in research.

The ML community has very strong norms around openness: free sharing of data, algorithms, models, and research papers. These strong openness norms appear to be broadly motivated by (1) distributing the benefits of research widely by making it accessible to all of society, and (2) enabling scientific progress by making it easier for researchers to critique and build on one another’s work.

Research practices that attempt to limit mal-use by being more cautious about how it is released and distributed necessarily reduce some forms of openness. Some who take openness to be a fundamental value in research may therefore disagree with such practices on principle. However, there are multiple different aspects to openness in research, and, as we’ve tried to highlight in this paper, multiple different approaches to being cautious about research release. Not all of these will necessarily be in tension with one another, and more exploration of research practices that decrease risk while protecting the most important aspects of openness would be valuable.

\subsection{Beliefs about risks}

Some disagree about the relative size of different risks involved in ML research. 

On the one hand, there is the risk that advances in ML might be misused by malicious actors in potentially catastrophic ways, which we’ve discussed. But restricting the release of ML research also creates its own risks: (1) of increasing power concentration, as a few research groups disproportionately control how ML capabilities evolve, and (2) of creating public confusion or even panic, by creating the impression that advances are more threatening than they are.

Beliefs about the relative size of these risks can lead to two very different perspectives. Those who believe that ML advances will lead to significant harm very soon may want to risk such power concentration in order to safeguard democracy and public trust in the long term. By contrast, for those who think weaponization is less immediately relevant and that we can reassess risks in the future, the costs of restricting research may seem less palatable.

While there is a genuine tension here, it is worth considering approaches that could address both sides of the concern (or at least address one side without exacerbating the other.) For example, some standardization of release practices, potentially managed by external entities, could help mitigate misuse without leading to power concentration.

\subsection{Beliefs about efficacy} \label{sec:efficacy}

Another dimension of disagreement centers not around what the risks are but how effective different practices are likely to be at reducing them.

Given strong incentives or low barriers to develop a technology (or achieve an insight), some suggest it is impossible to prevent either from leading to mal-use the long run, which could mean that restricting the release of research with potential for mal-use is futile. Others suggest that we can significantly impact incentives or barriers, or that slowing down release into the world can still make a significant difference, especially if this gives us time to build defenses against potential mal-use. There is also the perspective that it is easier to build systems to defend against mal-use if more research is public, and the counterview that public information can sometimes help attackers more than defenders (`security through obscurity' may be unnecessary for e.g. keeping data private but is still allegedly crucial for anti-spam defense). As ML researchers continue to experiment with release practices and explore similar challenges in other fields, we may learn about the efficacy of different approaches which can help inform these beliefs.

\subsection{Beliefs about future needs}

Finally, there’s a question of whether we might eventually need processes for release of ML research, even if they’re not essential now. 

For those who believe that we might develop much more advanced ML systems in the relatively near future, and that potential for harm will increase with these advances, then it probably makes sense to start developing careful norms and processes now regardless of current harms. For those who are more skeptical of the possibility of much more advanced capabilities, think that such capabilities are unlikely to be dangerous, and/or that restricting release is unlikely to be effective in the future regardless, developing such processes now looks unnecessary.

\begin{center}
\mbox{*}
\end{center}

Part of the reason for laying out various different options for release of research is to show that this needn’t be a polarized debate: it’s not a simple choice between `open' or `closed' ML research. It's worth considering whether, within our menu of options, there are approaches which can strike a balance between the differing perspectives outlined here.

\section{Recommendations}

We've laid out some considerations, tools, and options for thinking through release of potentially harmful research in a nuanced way. But what must be done now? Here are some brief recommendations:

\begin{enumerate}
    \item \textbf{Increase understanding} of the risk landscape and possible mitigation strategies:
    \begin{itemize}
        \item \textit{Develop standardized language} for talking about these issues e.g. around hazards, adversaries, mitigations and release options.
        \item \textit{Map risks} of different types of ML research in collaboration with subject matter experts, such as e.g. misinformation security researchers for synthetic media. Map out both immediate direct threats and potential longer-term path dependencies, in ways that address researcher concerns around risk hyperbole. Develop practices for safely discussing such risks.\footnote{This type of investigation might also be referred to as e.g. threat models, risk analysis, or impact analysis, each of which involves a different set of useful lenses.}
        \item \textit{Map mitigation options}, e.g. ways of reducing the harms resulting from mal-use of synthetic media research, and the stages/times at which they are applicable.
    \end{itemize}
    \item \textbf{Build a community and norms} around competency in understanding the impacts of ML research:
    \begin{itemize}
        \item \textit{Establish regular workshops} to focus on release challenges.
        \item \textit{Spread awareness} of the risks of ML research to both groups who might be affected and who can help mitigate the risks. Proactively seek to include and learn from those who have been impacted.
        \item \textit{Encourage impact evaluation} both positive and negative, for research publications, presentations, and proposals (such as that proposed by the ACM FCA \cite{hecht2018}).
    \end{itemize}
    \item \textbf{Fund institutions and systems} to grow and manage research practices in ML, including potentially:
        \begin{itemize}
            \item \textit{Support expert impact evaluation} of research proposals, so that the burden of this does not fall entirely on individual researchers (who may not have the relevant expertise to assess hazards). This might involve e.g. identifying groups with subject matter expertise who can do evaluations (at the request of researchers), coordinating, and potentially even paying for review.
            \item \textit{Prototype vetting systems} to help enable shared access to potentially sensitive research (as opposed to the current system where researchers attempt to validate if those requesting their models are malicious actors \cite{groverreview}, often via error-prone ad-hoc Googling).
            \item \textit{Develop release procedures} for research already deemed to raise potential risks (managing all of the above if needed, so that individual researchers can spend more time on actual research while still mitigating risks). Currently, organizations are unilaterally not publicly releasing results, so developing better procedures could actually open up research.
        \end{itemize}
\end{enumerate}

\section{Conclusion}

It is clear that advances in ML have the potential to be misused: the main example we have discussed here is how advances in synthetic media creation may be used to sow disinformation and mistrust (but many others can be, and have been discussed \cite{brundage2018malicious}). We must start thinking about how to responsibly safeguard ML research. 

Here we focus on the role of release and publication practices in preventing mal-use of ML research. The idea that we might sometimes restrict research release has been met with understandable concern from parts of the ML community for whom openness is an important value. Our aim here has been to decrease polarization in this debate; to emphasize that this is not a simple choice between ``open'' and ``closed'' ML research. There are a variety of options for how and when different aspects of research are released, including many drawn from parallels to existing fields, and many possible processes for making these decisions. 

There will always be disagreements about the relative risks and benefits of different types of research, the effectiveness of different mitigation strategies, and ultimately how to balance the values of openness vs. caution. We must more deeply explore the risks and options, and develop release strategies and processes that appropriately balance and manage the trade-offs.

Ultimately, we want research to benefit humanity. We see this work as part of a maturing of the ML community, alongside crucial efforts to ensure that ML systems are fair, transparent, and accountable. As ML reshapes our lives, researchers will continue to come to terms with their new powers and impacts on world affairs.


\bibliographystyle{unsrt}  
\bibliography{references}  



\emph{We are grateful to the many researchers across academia, industry, and civil society who provided invaluable feedback. This is meant to be just the start of a conversation, and we are likely to update this document as we learn more. If you have thoughts or feedback, or would like to otherwise contribute to this discussion, please reach out to the authors via email or on Twitter.}

\end{document}